# The insoluble problems of books: What does Altmetric.com have to offer?

## Daniel Torres-Salinas[1], Juan Gorraiz[2] and Nicolas Robinson-Garcia[3*]

[1] Universidad de Granada (EC3metrics & Medialab UGR), Granada (Spain)
[2] Vienna University Library, University of Vienna, Vienna (Austria)
[3] INGENIO (CSIC-UPV), Universitat Politècnica de València, Valencia (Spain)
* Corresponding author: elrobin@ingenio.upv.es

## Abstract

**Purpose:** Analyze the capabilities, functionalities and appropriateness of Altmetric.com as a data source for the bibliometric analysis of books in comparison to PlumX.

**Methodology:** We perform an exploratory analysis on the metrics the *Altmetric Explorer for Institutions* platform offers for books. We use two distinct datasets of books. On the one hand, we analyze the Book Collection included in Altmetric.com. On the other, we use Clarivate's Master Book List, to analyze Altmetric.com's capabilities to download and merge data with external databases. Finally, we compare our findings with those obtained in a previous study performed in PlumX.

**Findings:** Altmetric.com combines and orderly tracks a set of data sources combined by DOI identifiers to retrieve metadata from books, being Google Books its main provider. It also retrieves information from commercial publishers and from some Open Access initiatives, including those led by university libraries such as Harvard Library. We find issues with linkages between records and mentions or ISBN discrepancies. Furthermore, we find that automatic bots affect greatly Wikipedia mentions to books. Our comparison with PlumX suggests that none of these tools provide a complete picture of the social attention generated by books and are rather complementary than comparable tools.

**Practical implications:** This study targets different audiences which can benefit from our findings. First, bibliometricians and researchers who seek for alternative sources to develop bibliometric analyses of books, with a special focus on the Social Sciences and Humanities fields. Second, librarians and research managers who are the main clients to which these tools are directed. Third, Altmetric.com itself as well as other altmetric providers who might get a better understanding of the limitations users encounter and improve this promising tool.

**Originality/value:** This is the first study to analyze Altmetric.com's functionalities and capabilities for providing metric data for books and to compare results from this platform, with those obtained via PlumX.

# 1. Introduction

Many types of indicators have been suggested for the evaluation of books (Zuccala and Robinson-Garcia, forthcoming). Still, this publication type represents a weak spot when applying bibliometric techniques to assess the scientific impact of research. Most of the attention given to monographs is coming from bibliometricians specialized in the assessment of research in the Social Sciences and Humanities (SSH). This is partly due to the many changes taking place in the publishing market, in addition to the emergence of new databases, analytical tools and information providers. Moreover, there is an increasing demand from policy makers to develop metrics and indicators that can better assess the performance of SSH. In the last decade, all of this has led to a renaissance of studies devoted to this particular issue. An example of such interest is the launch of the "European Network for Research Evaluation in the Social Sciences" (http://enressh.eu). This initiative includes several working groups with one of them, 'Databases and uses of data for understanding SSH research', which, among other goals, aims at developing alternative metrics for the SSH.

Among the strands of SSH research devoted to the problem of books, there are two closely related to the present study. The first includes all studies devoted to the analysis and description of new databases and sources, and the second, though not mutually exclusive from the first, includes the construction and development of new indicators. This study explores and describes opportunities and limitations related to the use of Altmetric.com as a novel data provider for book metrics. The term "Altmetrics" was originally used to refer to metrics derived from social media activity and other alternative sources of information which go beyond the scientific realm (Priem *et al.*, 2010). But lately, the term seems to have become a "basket" concept, with very different metrics and sources included in the same mix (Wouters, Zahedi and Costas, forthcoming; Glänzel and Gorraiz, 2015). Their relevance is derived from the increasing interest they have raised as a potential means of capturing broader forms of impact (Bornmann, 2014). This is especially important in SSH research evaluation, where some scholars have attributed the limitations of bibliometrics to the fact that these fields target a broad array of audiences (Nederhof, 2006; Hammarfelt, 2014).

The enthusiasm surrounding altmetrics has created a window of opportunity for publishing firms, who have oversold the benefits of these new metrics, with the assistance of librarians, who "have become an important ally in the promotion of altmetrics, and a gap has opened between what librarians recommend researchers use and what researchers actually use" (Robinson-Garcia et al., 2017, p. 3). The present paper aims at bridging the gap between bibliometricians and academics, and information professionals and practitioners. We target especially university librarians and subject specialists who are in direct contact with the academic community and are direct consumers of tools for developing metrics. Our goal is to offer an overview and assessment of the new features offered by Altmetric.com directed at providing altmetric indicators for books. Our findings could be of interest for Altmetric.com or other altmetric aggregators, as many of the issues explored could help indicate how more reliable and robust data might be provided to various users and consumers.

The remainder of this paper is structured as follows. First, we will briefly review previous studies analyzing data sources for bibliometric purposes with a special focus on books, as well as different indicators proposed to assess their scholarly impact. The next section, will describe the information provided by Altmetric.com and the set of indicators and platforms it covers. Here we will focus on the Book Collection available through the Altmetric Explorer for Institutions, including a general overview of the information offered and sources from which the books are indexed. Section 4 presents a different approach with regard to the use of altmetric data. Given a set of books, altmetric indicators are retrieved and analyzed. Here we explore data retrieval and processing issues, coverage by fields and potential limitations. The following section discusses the pertinence, usability and reliability of Altmetric.com when referring to the analysis of monographs. While some of the issues discussed are common to other sources and tend to be related to the nature of books and certain conceptual problems (Torres-Salinas, Robinson-Garcia, et al., 2014), others are specific to Altmetric.com and should be taken into account if considering its use for the analysis of the impact of books. Section 6 compares the results provided and data retrieval issues with those found when using PlumX (Torres-Salinas, Gumpenberger and Gorraiz, 2017). We then conclude with some final remarks.

## 2. Antecedents

An avalanche of studies exploring new data sources took place at the beginning of the decade, with the launch of Web of Science's Book Citation Index (Leydesdorff and Felt, 2012; Gorraiz, Purnell and Glänzel, 2013; Torres-Salinas, Robinson-Garcia, et al., 2014), the inclusion of books in Scopus (Kousha, Thelwall and Rezaie, 2011) and the increase in research focused on Google Scholar (Kousha, Thelwall and Rezaie, 2011) and Google Books (Abdullah and Thelwall, 2014; Kousha and Thelwall, 2015a). They have also brought to light some of the characteristics and peculiarities (Torres-Salinas et al., 2012, 2013; Torres-Salinas, Robinson-García, et al., 2014; Chi, 2016; Glänzel, Thijs and Chi, 2016), which cannot be discerned on the basis of micro-analytic case studies or small samples of data.

More importantly, they show what difficulties need to be overcome with the indexing books and data that can be used for bibliometric purposes. These difficulties have to do with conceptual issues related to the nature of books, which are different from journal articles. Dealing with different editions of the same oeuvre, translations, etc., increases the difficulty of applying bibliometric solutions to measure the citation impact of these outputs. This is due to the intellectual and physical properties of books, which do not adhere to the same metadata structure used in journal article databases, leading some authors to suggest the treatment of books as "families of works" when applying bibliometric techniques (Zuccala et al., 2018).

These analyses have also confirm the limited capacity for citation analyses to address the impact that books can have on different types of audiences (Nederhof, 2006; Hammarfelt, 2014). To address this limitation, many have devoted their efforts to explore alternative indicators, such as library holdings or 'libcitations' (Torres-Salinas and Moed, 2009; White et al., 2009; White and Zuccala, 2018), book reviews (Zuccala and van Leeuwen, 2011; Gorraiz, Gumpenberger and Purnell, 2014), or publishers' rankings based on survey data (Giménez-Toledo and Román-Román, 2009; Giménez-Toledo, Tejada-Artigas and Mañana-Rodríguez, 2013). Most of these alternative indicators have not yet reached maturity. Yet, the interest in altmetrics has also caught up within the SSH research evaluation community (Giménez-Toledo, 2018). The growth of initiatives encouraging Open Access for monographs, such as the project OPERAS (Open access Publication in the European Research Area for Social Science and Humanities), increases the capability of altmetrics to track and capture the social attention of books. While the meaning of altmetrics may be well off from any notion of 'quality' or 'impact', it has been recently suggested that altmetrics can serve as "both drivers and outcomes of open science practices" (Wilsdon *et al.*, 2017, p. 11). This opens the possibility of exploring these metrics as a means for tracking the transition towards open science of monographs.

## 2.1 Studies on altmetrics for books

Recently, there have been some studies designed to explore the potential of altmetrics to capture the impact of books. Hammarfelt (2014) was the first, but at the time of the study, no altmetric provider covered monographs. This led him to search manually in different social media platforms for mentions to books by querying completely or partially book titles. What he found was that alternative metrics shared the same limitation as traditional bibliometrics which is the bias toward English language outputs. With the usage of books still shifting to electronic format, there can be yet another limitation to the application of altmetric indicators. Still, he noted a high frequency of tweets to books and concluded that altmetrics do hold promise for research assessments within the Humanities.

Other platforms, beyond those generally offered by altmetric providers, have also been explored (Kousha and Thelwall, 2015b, Zuccala et al., 2015). Zuccala et al. (2015) compared review ratings from the Goodreads platform with citations to history books (as cited in the Scopus journal literature). What they found was that books included in international libraries (defined as those included in the WorldCat union catalog) had a greater chance of being reviewed. Similar correlations were found with Amazon reviews (Kousha and Thelwall, 2016), leading the authors to conclude that online reviews tend to be related to a wider understanding of popularity, instead of academic impact.

Altmetric providers are now beginning to show up as a new data source for books. But the inclusion of offline indicators (e.g., policy briefs, library holdings) has led some to suggest that they are moving away from their original conception (Wouters, Zahedi and Costas, 2018). Originally, the term altmetrics emerged due to "the growth of new, online scholarly tools" which reflected "the broad, rapid impact of scholarship" (Priem et al., 2010). In terms of books, Torres-Salinas, Gumpenberger and Gorraiz (2017) recently found that social media platforms are not the most promising for developing indicators; rather it is usage counts, library holdings and reviews, with the library holding count proving to be the most prominent (Torres-Salinas, Robinson-Garcia and Gorraiz, 2017). Nevertheless, the book's transition to digital form would again limit the extent to which these underdeveloped indicators might be used.

## 2.2 Altmetric providers: Differences and discrepancies

Aside from indicators, there is also a critical need to understand how data providers harvest, aggregate and present data to their users. Indeed, some studies have been devoted to the analysis of specific altmetric providers (Robinson-García *et al.*, 2014; Zahedi, Costas and Wouters, 2014), as well as to comparisons between different providers (Zahedi, Fenner and Costas, 2014; Ortega, 2017; Peters *et al.*, 2017; Zahedi and Costas, 2018). Not only do we see differences concerning the set of metrics proposed by each provider, but more strikingly, differences in the information provided for the same indicators across different platforms.

These discrepancies have been reported in all comparative studies, but it is in the work of Zahedi and Costas (2018) where some explanations as to why this happens are made. Here we note three of the factors behind: time of retrieval, aggregation of data and data retrieval process. While the first factor seems reasonable, it does not explain why low correlations have been reported in a study between different providers offering the same metric (Zahedi & Costas, 2018). The most reasonable explanation seems to be that different decisions have been made in terms of the data presented, and how they have been collected. However, these studies have focused on journal articles, which have for many years now completed their shift towards electronic format and make use of identifiers (e.g., DOI, PMID), which certainly eases the retrieval process. Books present a more specific challenge, given that aside from ISBNs, there is no other useful identifier. Even in the case of ISBNs, a book may have different ISBNs by edition, format, translation and so on.

# 3. Altmetric.com Book Collection

In this section we will describe what can be found when consulting data directly from the *Altmetric Explorer for Institutions* web platform. We will start by giving an overview of raw numbers by type of output, and year. Then we will analyze data sources tracked by Altmetric.com to identify books and track mentions in social media.

## 3.1 General overview: research outputs, mentions and evolution

We consulted the *Altmetric Explorer* in May 2018 using the advanced search option in its main menu and analyzed total figures retrieved by type of output. With few exceptions (e.g., Peters *et al.*, 2016, 2017) most of the studies related to the use of altmetrics are based on the analysis of journal articles. However, Altmetric.com retrieves mentions also to datasets, books, book chapters, clinical trials and news stories. This latter output type refers to publications in popular science magazines (e.g., The Conversation), science dissemination web portals (e.g., ecancer.org), or more commonly, comments, perspective articles, policy forums, etc. published in Nature and Science.

Table 1 includes a general view of the contents indexed in Altmetric.com at the time of the query. Altmetric.com contains almost 21 million records indexed in its database, around half of them including at least a mention in any of the metrics it covers. Almost 70% of the outputs are journal articles, with books representing around 5% and book chapters 2.3%. Still, together, they do not even represent 1% of the outputs with mentions and even more negligible is the proportion of mentions directed at these outputs. Books have an average of more than two mentions by book, which increases up to more than 3 when removing those with no mentions at all. In the case of book chapters, it seems that altmetric data is of no relevance to assess their potential impact, as their average is near to zero mentions by book chapter.

Table 1. Contents of Altmetric.com by type of output

| Type of research output | Research Outputs | Outputs with mentions | Total mentions | Average Mentions |
|---|---|---|---|---|
| Article | 14,494,667 | 10,248,575 | 67,789,339 | 4.68 |
| Dataset | 26,888 | 21,206 | 112,959 | 4.20 |
| Books | 1,189,253 | 818,135 | 2,675,537 | 2.25 |
| Books Chapters | 5,044,984 | 78,959 | 2,238 | 0.00 |
| Clinical Trials | 36,882 | 35,646 | 17,298 | 0.47 |
| News Stories | 96,609 | 95,251 | 8,893,913 | 92.06 |
| **All outputs** | **20,889,929** | **11,298,415** | **79,890,012** | **3.82** |

Although the inclusion of books was announced in 2016 (Liu, 2017), Altmetric.com includes books dated from far before. This is not due to a conscious effort on tracing books back in time, but, as will be discussed later, due to the data retrieval process employed. This explains the lack of pattern when analyzing number of books indexed by publication year (see Table 2). While articles show an increasing and stable pattern, books do so to a lesser extent, and book chapters do not show any kind of pattern. Interestingly, this is less evident when focusing on books and chapters with mentions. Where we do find a positive growth, if an important increase in the last two years analyzed, is on the number of mentions directed at books and book chapters, which could be signaling a transition towards the electronic format, which facilitates sharing and capturing mentions to books. This is explained by an increase of Wikipedia mentions in the two last years.

Table 2. Longitudinal comparison between journal articles, books and book chapters for the 2013-2017 period

| Research outputs | 2013 | 2014 | 2015 | 2016 | 2017 |
|---|---|---|---|---|---|
| Articles | 935,802 | 972,635 | 1,103,626 | 1,259,912 | 1,227,340 |
| Books | 110,410 | 43,059 | 32,889 | 57,073 | 51,125 |
| Book Chapters | 300,970 | 299,669 | 240,548 | 270,995 | 254,509 |
| **Outputs with mentions** | **2013** | **2014** | **2015** | **2016** | **2017** |
| Articles | 694,263 | 750,063 | 896,855 | 1,056,389 | 1,042,993 |
| Books | 30,588 | 28,038 | 46,763 | 40,422 | 37,867 |
| Book Chapters | 4,237 | 5,069 | 9,639 | 16,644 | 11,150 |
| **Total Mentions** | **2013** | **2014** | **2015** | **2016** | **2017** |
| Articles | 4,496,193 | 5,455,146 | 7,578,038 | 10,140,826 | 12,703,226 |
| Books | 156,282 | 159,551 | 257,024 | 368,919 | 467,534 |
| Book Chapters | 22,623 | 14,017 | 26,347 | 47,550 | 38,223 |

Table 3. Coverage of Altmetric.com by platform for Book Collection mentioned in the 2000-2018 period

| Platform | Total records | Share |
|---|---|---|
| **Twitter** | 209682 | 26.2% |
| **Mendeley** | 162896 | 20.4% |
| **Syllabi** | 14,226 | 17.8% |
| **Wikipedia** | 93376 | 11.7% |
| **Facebook** | 40845 | 5.1% |
| **Blogs** | 40429 | 5.1% |
| **News media** | 36657 | 4.6% |
| **Policy** | 22556 | 2.8% |
| **Google Plus** | 10510 | 1.3% |
| **Q&A** | 3237 | 0.4% |
| **Patents** | 3185 | 0.4% |
| **Weibo** | 90 | 0.0% |
| **Peer review** | 84 | 0.0% |
| **F100** | 61 | 0.0% |
| **LinkedIn** | 53 | 0.0% |
| **Total** | **799098** | **100.0%** |

Finally, Table 3 shows coverage by social platform to books included in the Book Collection mentioned between 2000 and 2018. Similarly to journal literature, Twitter is the platform with the largest coverage, followed by Mendeley. Here we must note that Altmetric.com only retrieves data from Mendeley when a record is mentioned in any of the other platforms included (Robinson-García *et al.*, 2014). The most interesting finding is the importance of Wikipedia mentions which covers more than 10% of the Book Collection. This finding contrasts with the low coverage of Wikipedia found on journal literature (Zahedi, Costas and Wouters, 2014).

## 3.2 Information sources for tracking books

Altmetric providers make an extensive use of unique identifiers and URL linkages to track mentions in social media platforms. Indeed, their reliance on DOI numbers is widely known and has become a topic of concern when discussing their limitations (Robinson-García *et al.*, 2014; Zahedi and Costas, 2018). In the case of books, ISBNs do not provide traceable links as DOIs do. To overcome this issue, Altmetric.com seems to track a series of data sources from which they extract all book records whenever DOIs are not available. Table 4 shows the top web domains reported by Altmetric.com as points of access to book data. It refers to 799,098 books published between 2000 and 2018 included in the Book Collection. As observed, the reliance on DOIs is still quite high (41% of books) considering that its use is not as extended in monographs as it is with journal literature. Its tracking system is similar to that employed with journal literature (Zahedi & Costas, 2018, Table 11), systematically tracking different sources in an orderly manner. It first tracks the DOI number− explaining why 75% of books with DOI do not show information related to their point of access−, followed by Google Books and a set of publishers, along with other institutions. In this "other", we must highlight the inclusion of international and governmental organizations like the World Bank, the publications portal from the European Commission, OECD or the US National Academies of Sciences, Engineering and Medicine. Even corporations like RAND are included. These organizations author reports or policy documents which are, in many cases, presented in a monograph format and with an ISBN. This means that an important number of records indexed as books in Altmetric.com are actually reports and policy briefings which do not adhere to the traditional notion of an academic book.

In the case of books without a DOI number, Google Books is the main source used. 37% of books do not include an access point in their records. But still, it is possible to link to their description from the *Altmetric Explorer* platform. After randomly searching for sources for books without access point, we find that they come to a great extent from Harvard Library (http://library.harvard.edu), which has an Open Metadata policy, allowing third parties to track their entire catalog. This is observed with many other libraries as well as with Open Access initiatives for books like OAPEN (http://oapen.org) or the Swedish National Library (http://urn.kb.se).

Table 4. Top 18 points of access used by Altmetric.com to retrieve books mentioned in the 2000-2018 period, ranked by total number of records and distinguished by those with and without DOI number. In bold URLs to nonprofit and governmental organizations providing reports. In italics those linking to library catalogues and open portals of books.

| Without DOI (315 total sources) | 471,347 | With DOI (129 total sources) | 327,751 |
|---|---|---|---|
| books.google.com | 272,555 | No URL link | 244,797 |
| No URL link | 176,508 | books.google.com | 40,660 |
| market.android.com | 9,443 | link.springer.com | 17,347 |
| store.elsevier.com | 2,715 | **documents.worldbank.org** | **7,273** |
| **rand.org** | **1,989** | **nap.edu** | **6,154** |
| uk.sagepub.com | 1,385 | onlinelibrary.wiley.com | 1,803 |
| us.sagepub.com | 1,157 | **oecd-ilibrary.org** | **1,801** |
| berghahnbooks.com | 1,129 | **rand.org** | **1,398** |
| elsevier.com | 907 | tandfebooks.com | 1,336 |
| **publications.jrc.ec.europa.eu** | **504** | market.android.com | 867 |
| **adb.org** | **392** | **un-ilibrary.org** | **864** |
| *oapen.org* | *324* | elgaronline.com | 714 |
| **nap.edu** | **278** | booksandjournals.brillonline.com | 390 |
| thieme.de | 276 | **publications.jrc.ec.europa.eu** | **223** |
| **iwmi.cgiar.org** | **267** | *oapen.org* | *194* |
| elsevierhealth.com | 162 | uk.sagepub.com | 163 |
| ebookstore.thieme.com | 161 | *urn.kb.se* | *143* |
| facetpublishing.co.uk | 103 | www.elsevier.com | 138 |

# 4. Data retrieval and processing of altmetrics for a collection of SSH books

In this section, we test the capacity of Altmetric.com to retrieve data given a predefined set of books. This is a basic feature needed by any institution considering a subscription to this tool, where the aim is to monitor the altmetric impact of the book collection of their institution. To perform this test, we downloaded all books included in the Master Book List from Clarivate's Book Citation Index (http://wokinfo.com/mbl/). A total of 60,239 books were retrieved from this list, including book title, ISBN, publisher, subject category and series title. We used the ISBN data to query the *Altmetric Explorer*. It is noteworthy that the Master Book List provides a single ISBN by book, which means that there may already be a loss of information if a book has more than one ISBN identifier.

A total of 33,014 books were retrieved from Altmetric.com and downloaded in order to match altmetric mentions with our original table. Only 15,545 records were successfully matched back to our original dataset (47% of the retrieved set and 26% of the original set). After carefully looking for potential discrepancies on format, etc., we find out that this is due to ISBN mismatches. Although we queried for a single ISBN code by record, Altmetric.com retrieves mentions to all ISBN codes identified that belong to the same book. The list of ISBN codes for each book are actually displayed in the web platform. Still, when downloading the data from Altmetric.com, only one ISBN by record is offered. This is what makes it almost impossible to automatically match these retrieved records with our original dataset.

Figure 1. Number of books with mentions in Altmetric.com and matched and number of mentions received by subject category.

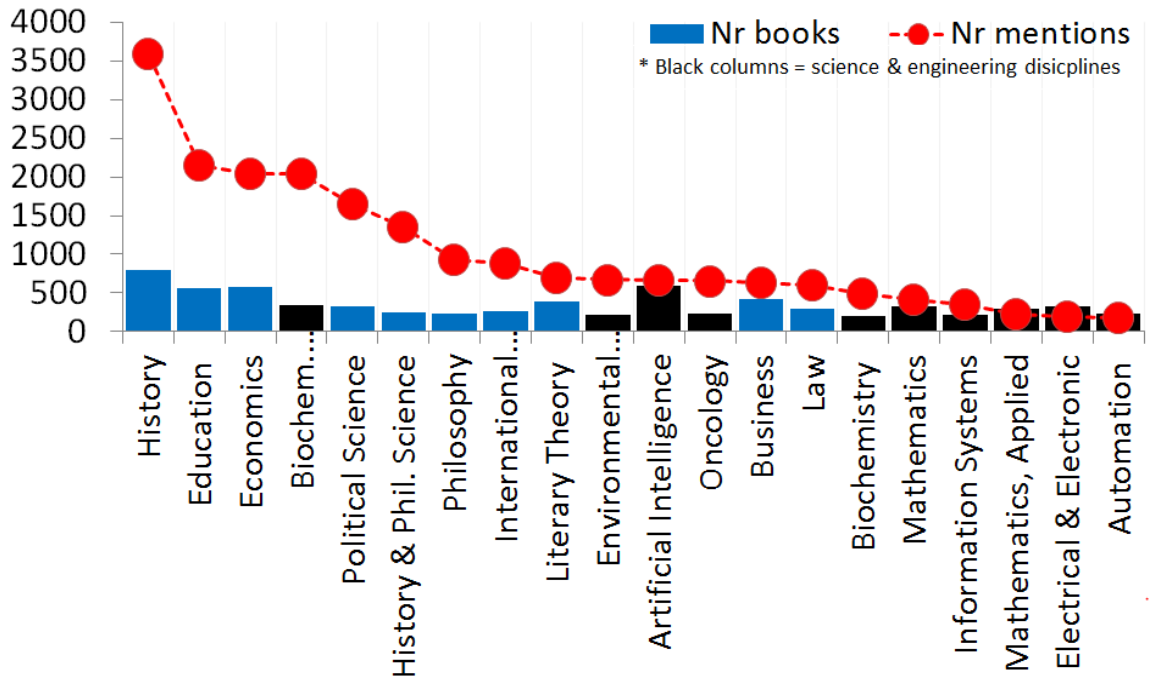

Table 5. General overview of books by SSH subject category in Clarivate's Master Book List and records and mentions retrieved and matched from Altmetric.com

| Discipline | Books in Web of Science | Total mentions | Research outputs | % Research outputs | Outputs with mentions | % Outputs with mentions |
|---|---|---|---|---|---|---|
| History | 4018 | 4355 | 2140 | 53% | 1434 | 36% |
| Literary Theory & Criticism | 2840 | 1010 | 1635 | 58% | 669 | 24% |
| Economics | 2810 | 4502 | 1775 | 63% | 719 | 26% |
| Education & Educational Research | 2723 | 2679 | 1401 | 51% | 541 | 20% |
| Political Science | 1865 | 2039 | 1213 | 65% | 649 | 35% |
| International Relations | 1723 | 1127 | 1003 | 58% | 508 | 29% |
| Business | 1665 | 1050 | 981 | 59% | 325 | 20% |
| Law | 1446 | 1480 | 728 | 50% | 438 | 30% |
| Philosophy | 1369 | 1259 | 739 | 54% | 371 | 27% |
| Religion | 1239 | 718 | 465 | 38% | 309 | 25% |

Even though we were only able to process and link back almost half of the retrieved books with mentions from Altmetric.com, we performed a descriptive analysis by subject category to offer an overview on potential differences by fields. We must stress that this is done only with those which we could match back to data downloaded from Clarivate's Master Book List, as any kind of systematic bias on this loss of information cannot be assessed in this analysis. Figure 1 and Table 5 display raw numbers and proportions of mentions and records analyzed based on our original data set. History is the subject category which comprises the largest number of mentions and outputs with mentions, followed by Education and Economics. In terms of coverage, Political Science and Economics are the categories with a largest share of books retrieved from Altmetric.com. History and Political Science are, out of the total of books indexed in Altmetric.com, the ones with the highest share of books with at least one mention in at least one of the platforms covered by Altmetric.com.

# 5. Main characteristics, strengths and limitations

This paper aims at exploring Altmetric.com as a source for retrieving altmetric indicators for books. The two previous sections were devoted to describing the tool and its usability. Throughout the text we have briefly mentioned issues that limit to a great extent the utility of this data source. In this section we will expand on significant issues observed when testing this tool that should be considered. To do so, we have structured this section into three parts, containing crucial aspects to consider when assessing a new data source: pertinence, usability and reliability.

## 5.1 Pertinence

The first issue to consider is the pertinence of Altmetric.com in particular and altmetrics generally, as indicators aligned with the phenomenon we intend to measure. This question is more of a conceptual rather than technical or methodological nature. Although such assessment falls out of the scope of this paper, this is an unavoidable question. The capability of altmetric indicators to measure broad forms of impact has long been contested (Sugimoto *et al.*, 2017). But still, altmetric indicators have the potential to provide unique data on other aspects of the scholarly communication system. Books and book chapters are a format which have not entirely transitioned to the digital format. While this may entail a shortcoming when aiming at monitoring their reception, altmetric indicators can be used to study the advancement of Open Access in monographs (Wilsdon *et al.*, 2017, p. 10). Since there is evidence that books are tweeted (Hammarfelt, 2014), institutions and publishers can use this information to identify topics which are best received by the audience or analyze the reception or attention their monographs receive.

We do find that book chapters tend not to be mentioned in social media and therefore the pertinence of Altmetric.com as a source to monitor social media attention of book chapters should be dismissed.

## 5.2 Usability

Due to the high reliance altmetric providers have on DOI codes, one of the key questions when analyzing Altmetric.com for books was to see how it deals with publication forms which cannot be unequivocally identified. Altmetric.com relies to a large extent on the use of DOI and ISBN identifiers, but not all records indexed in the book collection have identifiers. Regarding books with a DOI identifier, versus those without one, we find an unexpected pattern in the number of books indexed by publication year in the 2000-2018 period. There is a converging trend with a sustained increase of books with a DOI until 2012. Between 2012 and 2013 there is an abrupt increase of books with a DOI, followed by a lower increase on books without a DOI between 2014 and 2017 (see Figure 2). We were not able to trace the reasons behind these patterns, probably due to the addition of new sources of information from which to index books.

Figure 2.Evolution of A) proportion and B) average of the Altmetric Attention Score of books indexed with and without DOI

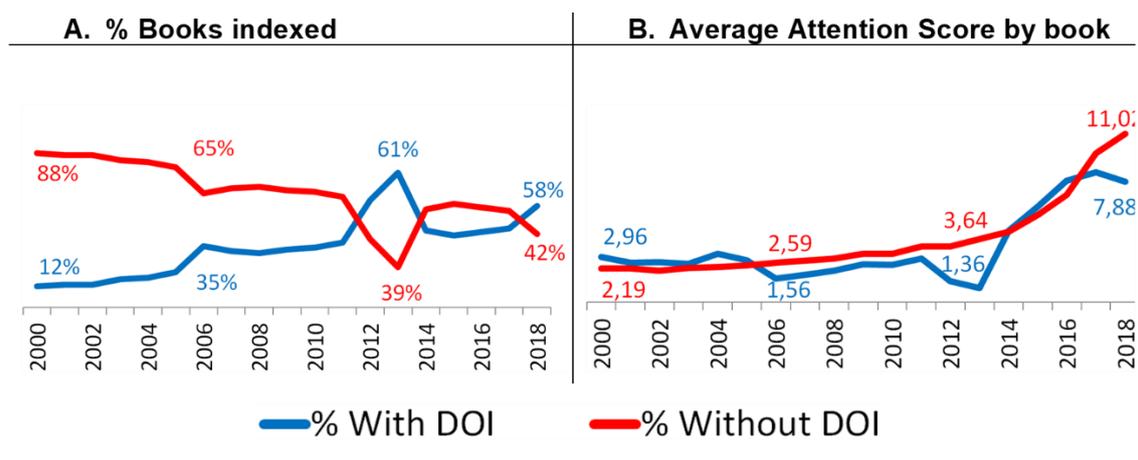

ISBNs are the other main identifiers used to link monographs. In this sense, Altmetric.com tracks book's description from Google Books as a first choice, and secondly, on data from libraries with an Open Metadata policy (e.g., Harvard University Library) or Open Access initiatives for books (e.g., OAPEN). All ISBN codes identified and belonging to a single book are merged. Still, some errors were identified in this merging process. For instance, we identified several cases where metrics attributed to one book pointed to another one. In other cases, there were empty records where the metadata information was not properly linked but actually had mentions pointing to 'real' books.

The main shortcoming attached to usability, was related to the incompleteness of data when downloading records from the *Altmetric Explorer*. As a data source used at the institutional level, it is expected to link the data provided with additional datasets and internal databases. For this, it is essential to have a field that can be matched univocally with the different datasets. The first choice when referring to monographs is to use the ISBN code of books which is also used to query Altmetric.com. However, Altmetric.com does not allow downloading the complete list of ISBN codes from each record but a single one. This means that the retrieved ISBN is not always the one that is used when querying Altmetric.com, which makes it impossible to link the data retrieved from external databases. By including all ISBN codes associated to each record in the data available to download, this would be easily overcome.

## 5.3 Reliability
The last essential aspect to consider is the reliability of the information provided. In this sense, and despite the work on merging ISBN codes pointing to a single book, we find for some books that many editions and re-editions are missing. Also duplicates can be found with different editions of a same book split into two different records. This is a common shortcoming when processing book data for bibliometric purposes.

Additional issues were identified, which led us to question to some extent the information provided by Altmetric.com. The first relates to the inflation of monographs caused by the inclusion of publication types that cannot be considered to be academic books. The second has to do with anomalies on the reported number of mentions received, where one case related to unreported changes in Altmetric.com and another related to the exogenous patterns of the platforms included. With regard to the inflation of publications, and as pointed out before, not all books included in the Altmetric.com Book Collection have a ISBN code assigned to them. From the total of books mentioned in the 2000-2018 period, 11,243 did not include an ISBN code in their record. This is not a significant number and only accounts to 1.4% of books. Still, we note that out of this share, half of them were retrieved from the World Bank (5,564 records) and almost 7% from RAND Corporation (770 records).

Regarding anomalies found on the reported number of mentions received by platform, we focus on two specific altmetric sources: Wikipedia and Syllabi. In the case of Wikipedia, we noted that almost 12% of monographs indexed in Altmetric.com had been cited in this platform (see Table 3). This coverage is outstanding when considering that for journal articles it is of roughly 1.5% (Zahedi, Costas and Wouters, 2014, p. 1495). Indeed, 47% of all mentions made from Wikipedia in 2017 to all outputs in Altmetric.com are directed to books (see Figure 3).

To analyze if this is due to a greater capability of Wikipedia to track books' influence or to another exogenous factor, we tracked which books had been cited in 2017. We observed that they all belonged to a series of books from the 19$^{th}$ Century which have been recently digitalized. These books where published between 1862 and 1873 to be precise and are books from the field of biology, which serve as catalogues of species. Indeed in May 2018 (date of the data retrieval), a total of 13,645 mentions from Wikipedia to four of these books were identified. These books are:

- Classification of the Coleoptera of North America prepared for the Smithsonian Institution by John L. LeConte. *4276 mentions*
- American Insects. *4033 mentions*
- Heteroptera, or true bugs of eastern North America, with especial reference to the faunas of Indiana and Florida, by W.S. Blatchley. *851 mentions*
- Catalogue of the specimens of heteropterous-Hemiptera in the collection of the British museum. By Francis Walker. *851 mentions*

Figure 3. Number of mentions from Wikipedia to A) books and articles between 2014 and May 2018, and B) Distribution of mentions by month between April 2017 and May 2018.

**A**

| Year of mention | Total mentions in Wikipedia | Total mentions to journal articles | | Total mentions to books | |
|---|---|---|---|---|---|
| 2014 | 107318 | 104923 | 98% | 965 | 1% |
| 2015 | 116036 | 111741 | 96% | 1905 | 2% |
| 2016 | 161276 | 155829 | 97% | 2672 | 2% |
| 2017 | 321464 | 165484 | 51% | 151563 | 47% |
| 2018 | 211137 | 123262 | 58% | 85142 | 40% |

**B**

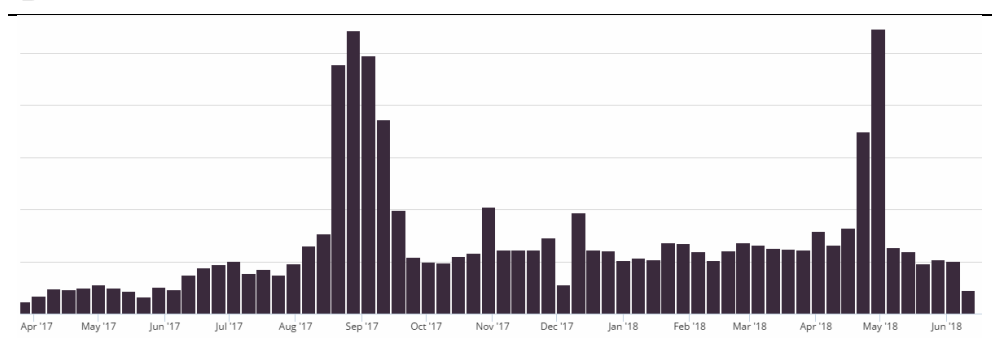

This figure reveals to some extent, the capacity of altmetrics to reflect the transition to electronic format of monographs, which can be of special interest to university libraries. But, in the case presented here, we identified that a single Wikipedia user was behind this massive number of mentions to 19$^{th}$ Century books: an automatic bot named Qbugbot, with even its own Wikipedia entry (see Figure 4). Indeed a superficial search tells us that it is not the only bot populating Wikipedia, which means that Wikipedia citations are greatly compromised if we do not know what they are really reflecting.

An anomaly of a different nature is the one related to syllabi mentions. Syllabi mentions are the only metric covered by Altmetric.com which specifically targets books. Its inclusion was announced in September, 2016, with information pointing to a new partnership, where syllabi were extracted from the Open Syllabus Project (Konkiel, 2016). This project seems to have been discontinued; although we cannot confirm this. As shown in Figure 5, we can only present an observed drop since 2003 onwards with almost no mentions to books from 2015 onwards.

Figure 4. Screenshot to the Wikipedia entry to Qbugbot available at https://en.wikipedia.org/wiki/User:Qbugbot

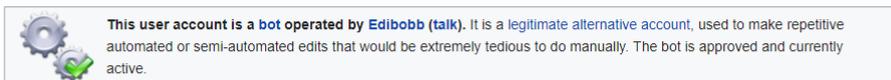

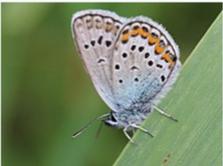

Figure 5. Distribution of mentions to books from syllabi in the 2000-2018 period.

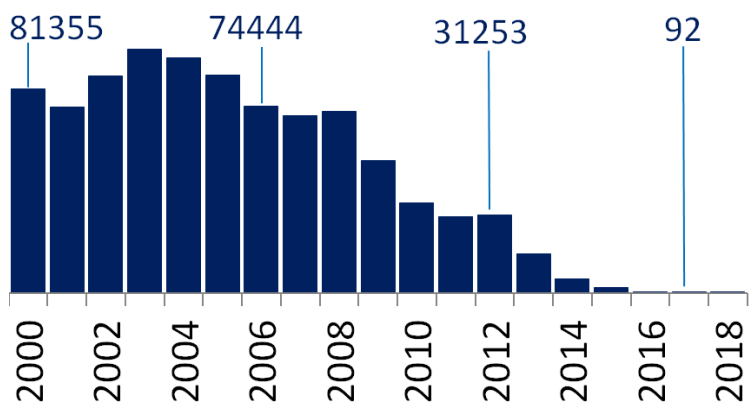

In this case, the anomaly relates to unreported changes in the sources tracked by Altmetric.com rather than with anomalous patterns of such sources. The fact that Altmetric.com is tracking a traditionally problematic publication output (in bibliometric terms) with metrics which are still underdeveloped, does not rule out entirely their potential use to track and monitor books' reception or impact. But the number of considerations and limitations attached to these types of assessments clearly increase, if not multiply.

# 6. Differences between Altmetric.com and PlumX

This paper aims at providing a full scope of the usability and potential of Almetric.com as a source from which to extract metrics for books. We do not address directly the potential of altmetrics for books but rather describe and analyze this tool. It is also not part of the scope of this paper to make comparisons with other altmetric providers which also offer metrics for books. Nevertheless, we can use findings from our previous research and experience with the provider PlumX. First of all, we must emphasize that none of the tools provide clear documentation on how the input data are exactly processed and how books are identified and tracked. It is almost a trial and error process to fully understand how they work and what it is exactly to fully understand how they work and what exactly they are providing to the user.

Both, PlumX and Altmetric.com, differ in terms of their philosophy and conceptual approaches to indicators. While Altmetric.com focuses on a single multi-indicator (the Altmetric attention

score[1]), PlumX has adopted a multi-approach based on five categories or dimensions (usage, citations, captures, mentions and social media), and presents itself as the tool to trace not only "altmetric" data but "all metrics".

In terms of data retrieval, in a previous study (Torres-Salinas, Gumpenberger and Gorraiz, 2017) we suggested that PlumX creates its own index of books. Once introduced it may enhance and aggregate automatically all bibliographic records with different variations of ISBN codes assigned and matched to external databases. Thus, different variations of ISBN codes are assigned and data matching with external databases seems to be much more smooth. However, neither provider can guarantee a complete data retrieval process. Arduous manual work will be required in order to exclude duplicates and correct mismatches. On the other hand, both tools support a minimum of transparency, with the retrieval of mentions, and the possibility of tracing back to an original source.

The data types included by each altmetric provider vary extensively, but for the most part, they tend to be rather complementary than comparable (Peters *et al.*, 2017). Also, PlumX covers a range of indicators from sources that seem to be much more relevant to books (i.e., online book reviews and library holdings ) than those covered by Altmetric.com. In this way, and especially in terms of library holdings, the PlumX platform has potential to overcome problems related to a lack of books in electronic format (Torres-Salinas, Robinson-Garcia and Gorraiz, 2017).

With regard to monographs, the difference approaches taken by each provider – i.e., the composite indicator versus multidimensional approach - means that PlumX includes variables that are not altmetrics (e.g., citations, usage). It also means that there are some social media platforms which are missing in this tool (see Table 6), however the coverage of these platforms in Altmetric.com (e.g., F1000, LinkedIn, Weibo) is near to zero. PlumX encompasses a total of 39 indicators versus the 18 indicators offered by Altmetric.com. But 21 of these are related to the Usage dimension (e.g., Repec, SSRN, E-print) which might be more pertinent for journal articles than books.

However, there are noteworthy inclusions in PlumX, which are of great interest when analysing the academic impact or reception of books, such as library holdings or reviews and scores in Amazon or Goodreads. In the case of library holdings, the coverage rate is extremely high up to 97% (Torres-Salinas, Gumpenberger and Gorraiz, 2017).

Table 6. Comparison between Altmetric.com and PlumX of coverage, mean of mentions per record and mean of mentions per indexed record for sources included in Altmetric.com. Data from PlumX is retrieved from table 1 of the supplementary material from Torres-Salinas, Gumpenberger and Gorraiz (2017). Note that book datasets differ.

| Altmetric.com | | | | | PlumX | | | |
|---|---|---|---|---|---|---|---|---|
| Platform | Coverage | Mean | Mean Available | | PlumX | Coverage | Mean | Mean Available |
| Twitter mentions | 26.2% | 2.28 | 8.70 | | Tweets:Twitter | 0,43% | 0.04 | 10.27 |
| Wikipedia | 11.7% | 0.22 | 1.95 | | Links:Wikipedia | 16.57% | 0.46 | 2.77 |
| News media | 4.6% | 0.15 | 3.45 | | News Mentions:News | 0.02% | 0.00 | 2,12 |

| Syllabi mentions | 17.% | 1.14 | 6.41 | Not included | --- | -- | -- |
|---|---|---|---|---|---|---|---|
| Blogs | 5.1% | 0.10 | 2.10 | Blog Mentions:Blog | 0.06% | 0.00 | 2.38 |
| Facebook | 5.1% | 0.10 | 2.04 | Facebook | 0.54% | 0.17 | 31.36 |
| Mendeley readers | 20.4% | 4.2 | 20.05 | Readers:Mendeley | 25.9% | 1,60 | 6.65 |
| Policy | 2.8% | 0.05 | 2.01 | Not included | --- | --- | --- |
| Google Plus | 1.3% | 0.02 | 1.82 | +1s:Google+ | 0,12% | 0.01 | 5.06 |
| F1000 | 0.0% | 0.00 | 1.08 | Not included | --- | --- | --- |
| LinkedIn mentions | 0.0% | 0.00 | 1.09 | Not included | --- | --- | --- |
| Patents | 0.4% | 0.00 | 2.03 | Not included | --- | --- | --- |
| Peer review | 0.0% | 0.00 | 1.83 | Not included | --- | --- | --- |
| Pinterest mentions | 0.0% | 0.00 | 1.08 | Not included | --- | --- | --- |
| Q&A | 0.4% | 0.00 | 1.15 | Not included | --- | --- | --- |
| Reddit mentions | 0.0% | 0.00 | 1.16 | Scores:Reddit | 0.01% | --- | 56.10 |
| Video mentions | 0.0% | 0.00 | 1.61 | Likes:YouTube | 0.00% | --- | 2.00 |
| Weibo | 0.0% | 0.00 | 2.86 | Not included | --- | --- | --- |

# 7. Concluding remarks

This paper tests the reliability and usability of metric indicators for books offered by Altmetric.com. In 2017, Altmetric.com announced that their *Altmetric Explorer for Institutions* now included metrics for monographs (Engineering, 2017). As noted elsewhere (Hammarfelt, 2014), books are still far from finalizing a shift from print to electronic format, which journal articles have. Altmetrics have the potential to trace the social uptake of research (Robinson-Garcia, van Leeuwen and Rafols, 2018) and can help to monitor the transition of scholarly outputs to Open Access (Wilsdon *et al.*, 2017). Furthermore, their speed and immediacy present them as 'catchy' indicators in the eyes of research managers, publishers and authors. This is especially relevant in evaluation of SSH research where citation windows are longer and analyses based on citations are limited.

This has not gone unnoticed by altmetric providers, and now PlumX and Altmetric.com offer such metrics and sell them to institutions and publishers. Although it was not part of the scope of this research to confirm the validity of using altmetrics in an evaluation context, it is still useful to determine exactly what types of information the altmetric providers are offering and how reliable it is. Previous studies have already highlighted and explained important differences concerning metrics reported for journal articles (Zahedi and Costas, 2018). Expanding these platforms in order to include monographs can only raise more questions about the selection of indicators provided, or features available to and essential for institutions subscribing to them. Issues such as the selection of indicators provided or the features available to users are essential for institutions considering subscribing to such products.

As a result of this study, which provides an overview of the Book Collection of Altmetric.com and its capabilities in terms of connecting with external databases, we can confirm that the product is still at an early stage. Moreover, many of the obstacles related to the indexing of monographs and problems encountered with other book databases only become more complex when intertwined with the volatile nature of altmetrics. For example, the deduplication of different monograph editions or translations can easily lead to incorrect altmetric assignments. Nevertheless, some issues are more easily addressed than others; thus platforms designed to make it possible to download a book's complete list of ISBN codes might eventually make it easier to link to external databases, like Current Research Information Systems (CRIS).

By exploring the main sources (e.g., Google Books; Open Access platforms and library catalogs) that Altmetric.com uses to retrieve information from books, we demonstrate the vital work that librarians do in order to support the development of these products, and make them more feasible.

Also, by comparing Altmetric.com to PlumX, and previous studies based on PlumX (Torres-Salinas, Gumpenberger and Gorraiz, 2017; Torres-Salinas, Robinson-Garcia and Gorraiz, 2017), we were able to conclude that, similar to the work of Peters et al. (2017), the two platforms are complementary rather than similar. With each platform we have observed different sets of indicators, which require different types of data retrieval and processing strategies. But, more importantly, we have found that there is a higher number of book-specific indicators (i.e., library holdings and Amazon/Goodreads review scores) in PlumX than in Altmetric.com.

And finally, we wish to emphasize that more documentation is needed at Altmetric.com so that users might better understand certain patterns behind the metrics they offer. We have identified two anomalies with regard to Wikipedia and syllabi mentions. In the first case, it is related to the existence of bots which automatically generate Wikipedia content and include references to scholarly work. The activity of one single bot in 2017 generated 47% of all Wikipedia mentions and that these were all directed to recently digitalized books from the 19[th] Century. In the second case, the anomaly resulted from what seems to be a discontinued project, as syllabi mentions seemed present in the database until 2014 when they ceased. Learning about these issues is crucial for any user who might use the information reported by Altmetric.com for any type of decision, as it will lead them to misinterpretations.

## Acknowledgements

The authors would like to thank Altmetric.com for providing access to their data through the Altmetric Explorer for Institutions. We acknowledge two anonymous reviewers and the guest editor for their comments and suggestions. Nicolas Robinson-Garcia is currently supported by a Juan de la Cierva-Incorporación grant from the Spanish Ministry of Science, Innovation and Universities.